\documentclass[showpacs,twocolumn,superscriptaddress,nofootinbib]{revtex4}

\usepackage{doi}
\usepackage{hyperref}
\hypersetup{
  colorlinks=true,        %
  linkcolor=blue,         %
  citecolor=cyan,         %
}

\usepackage{graphicx}
\usepackage{dcolumn}
\usepackage{bm}
\usepackage{color}
\usepackage{footmisc}

\usepackage{amsmath}
\usepackage{amssymb}


\begin{document}

\title{On overspinning of black holes in higher dimensions }

\author{Sanjar Shaymatov}
\email{sanjar@astrin.uz}

\affiliation{Institute for Theoretical Physics and Cosmology, Zheijiang University of Technology, Hangzhou 310023, China}\affiliation{Ulugh Beg Astronomical Institute, Astronomicheskaya 33, Tashkent 100052, Uzbekistan}
\affiliation{National University of Uzbekistan, Tashkent 100174, Uzbekistan}
\affiliation{Tashkent Institute of Irrigation and Agricultural Mechanization Engineers,\\ Kori Niyoziy 39, Tashkent 100000, Uzbekistan}

\author{Naresh Dadhich}
\email{nkd@iucaa.in}

\affiliation{Inter University Centre for Astronomy \&
Astrophysics, Post Bag 4, Pune 411007, India }

\begin{abstract}
It turns out that repulsive effect due to rotation of a rotating black hole dominates over attraction due to mass for large $r$ in dimensions $>5$. This gives rise to a remarkable result that black hole in these higher dimensions in contrast to lower dimensional ones cannot be overspun even under linear test particle accretion. Further if a black hole in dimension $>4$ has one of its rotation parameters zero, it has only one horizon and hence it can never be overspun. Thus rotating black holes in six and higher dimensions, and those with one rotation parameter switched off, would always obey the weak cosmic censorship conjecture (WCCC).

\end{abstract}
 \maketitle

\section{Introduction}
\label{introduction}

{The discovery of gravitational waves as that of two stellar black hole mergers \cite{Abbott16a,Abbott16b} through LIGO-VIRGO detection has opened a qualitatively new stage to the black hole astrophysics. Gravitational wave is expected to be a very powerful tool in revealing black hole's hidden properties that have remained unknown so far. After gravitational wave, very recently, the first image of the supermassive black hole in the center of M87 galaxy was obtained
under Collaboration of the Event Horizon Telescope (EHT)
\cite{Akiyama19L1,Akiyama19L6}. Testing the rotational nature of M87 galaxy from the circularity and size of its first image has also been very recently addressed~\cite{Bambi19}. This first image of black hole candidate also opens new prospect to probe black hole candidates more vigorously and definitively. However, there are yet many unexplored aspects of black hole. One, and perhaps the most important, among them is the Cosmic Censorship Conjecture (CCC) proposed by Penrose \cite{Penrose69} in 1969 which still remains an open and challenging question. Its validity strongly supports the presence of black holes with an event horizon, thereby concealing central singularity from observers outside. Even so it is still unproven, yet testing the CCC with various tools and physical processes allowing transition from black hole to naked singularity have remained an active area of  research. To test CCC a gedanken experiment is proposed to destroy black hole horizon by impinging test particles of appropriate parameters onto black hole and see whether cover of horizon is broken exposing the singularity. This has been one of the favourite topics with relativists for over three decades now. Is the end state of gravitational collapse always black hole or naked singularity~\cite[see,
e.g.][]{Christodoulou86,Joshi93,Joshi15,Joshi00,Goswami06,Harada02,Stuchlik12a,
Vieira14,Stuchlik14,Giacomazzo-Rezzolla11}]? However the question still remains open and braying for answer. The mute question in general relativity is, does collapse end up in black hole or naked singularity? In the latter case it is very important and exciting that the seat of infinitely large curvature would be exposed to physical enquiry. }

{Converting a black hole into a naked singularity by overcharging/spinning through test particles of appropriate charge and rotation parameters impinging on black hole has a long history. In a gedanken experiment~\cite{Wald74b}, it was envisaged that an extremal black hole is bombarded with test particles of suitable parameters so as to overcharge/spun it into a naked singularity. It was shown that extremal black hole horizon can never be destroyed and CCC holds good. It was to expose singularity to external observer, this would have been violation of CCC in the weak form --- WCCC. Much later it was shown~\cite{Dadhich97} that a non-extremal black hole cannot be converted into an extremal one by test particle accretion. This was all done for test particles impinging on via geodesics or Lorentz force trajectories. This however kept a question open, though extremality may not be attainable yet however it may be possible to be jumped over in a discontinuous process --- going over without pasing through extremal state, from non-extremal to over-extremal state in a discontinuous discrete process .}

The question is, is a transition from near-extremality to over-extremality achievable for a black hole in a discontinuous non-adiabatic process? This led to a new stage for probing the issue afresh. This experiment was first considered \cite{Hubeny99}, and it turned out that overcharging of black hole was indeed possible. It was later extended to overspinning of a rotating black hole \cite{Jacobson09}. Note that overcharging/spinning was initiated for a linear order particle accretion in which all higher order effects were not taken into account. This led to a spurt of activity as evidenced in \cite[see, e.g.][]{Hubeny99,Jacobson09,Saa11,Matsas07,Berti09,Shaymatov15,
Bouhmadi-Lopez10,Doukas11,Rocha14,Li13,Jana18,Song18,Duztas18,
Duztas-Jamil18b,Duztas-Jamil20,Yang20a}
exploring various situations and scenarios for overcharging/spinning of black hole\footnote{In particular in Refs.~\cite{Bouhmadi-Lopez10} and \cite{Doukas11}, overspinning of higher dimensional extremal black hole was considered by throwing in test particle geodesics in accordance with Ref.~\cite{Wald74b} and it was shown that it was not possible to create a naked singularity violating WCCC. In the former various rotating geometries were investigated and they all accorded to WCCC while  the latter had  studied Myers-Perry black hole in dimensions $D<10$. }.

{Later the above experiment was addressed by taking into account self-force and backreaction effects, and then it turned out that impinging particles won't be able to reach the horizon, thereby over-extremality could not be attained~\cite{Barausse10,Zimmerman13,Rocha11,Isoyama11,Colleoni15a,
Colleoni15b}. Inclusion of backreaction effects was also considered for a regular black hole~\cite{Li13}. These extensive works verify weak cosmic censorship conjecture when self-force and backreaction effects are taken into account. }

{In recent years much attention has been devoted to destruction of event horizon of black holes in various context and framework, for example massive complex scalar test fields around a black hole~\cite[see,
e.g.][]{Semiz11,Toth12,Duztas13,Duztas14,Semiz15,Shaymatov20d}, rotating anti-de Sitter black
holes~\cite{Gwak16,Natario16,Natario20}, BTZ black holes and fields~\cite{Shaymatov15,Duztas16}, magnetized black holes~\cite{Siahaan16,Shaymatov19b}, black hole with charged scalar field \cite{Gwak20}, and black hole in 4D EGB gravity~\cite{Yang20b}. Further WCCC has also been verified by considering black hole dynamics~\cite{Mishra19,Bardeen73b}. Also has been studied the phenomenon of spin precession for rotating black hole with a view to distinguish between black hole and naked singularity \cite{Chakraborty17}.} Not only that
if a naked singularity can be formed as a result of collapse, could it be converted into a black hole, has also been recently addressed~\cite{Hioki19}.

{Much of the previous analyses involved linear order accretion, very recently Wald and Sorce \cite{Sorce-Wald17,Wald18} have proposed a new gadanken experiment which includes second order particle accretion process. Then it turns out that black hole can never be over-extremalized and thereby WCCC is always obeyed. This has put to rest all linear order violations of WCCC, and further it has opened a new vista for study of non-linear accretion in various conditions and situations. Thus WCCC though may be violated at linear order, it would always be restored at non-linear order, see for example~\cite{An18,Gwak18a,Ge18,Ning19,Yan-Li19,Jiang20plb,Shaymatov19b}.

A recent analysis shows that five dimensional rotating black hole has a remarkable feature that it could be overspun under linear accretion when it has two rotations but not when it has only one rotation~\cite{Shaymatov19a}. Overcharging of a higher dimensional charged black hole has been studied~\cite{Revelar-Vega17}. This led to an interesting question - could black hole be over-extremalized when it has both charge and spin. There however exists no exact solution for an analogue of Kerr-Newman charged rotating black hole in five or higher dimensions. The only way to address this question is to consider the minimally gauged supergravity charged rotating black hole \cite{Chong05} exact solution in five dimensions. It is indeed closest to the Kerr-Newman black hole as it has all the expected and desired features for a charged rotating black hole in five dimension. What emerges here is that the ultimate result depends on which parameter, charge or rotation, is dominant. It is demonstrated that black hole with single rotation cannot be over-extremalized when rotation is dominant over charge while the opposite is the result when charge dominates over rotation~\cite{Shaymatov19c}.

{As dimension increases, number of rotations a black hole can have, also increases and it is given by $n = [(D-1)/2]$; i.e. $n=1, 2$ for $D=3,4$ and $D=5, 6$ respectively. Also we know that gravitational potential due to mass goes as $1/r^{D-3}$ which becomes sharper and sharper for increasing $D$ while that due to rotation goes as $1/r^2$ and higher orders. In $D\leq5$, attractive contribution due to mass would be able to dominate over the leading order repulsive rotation contribution\footnote{In $D=5$ the leading order term would be $-(\mu-a^2-b^2)/r^2$ where $\mu>a^2+b^2$ is required for existence of horizon. Here $\mu, a, b$ respectively refer to mass and two rotations of the black hole.}. That means five dimension is the threshold dimension where attractive component would be dominant everywhere outside horizon. In $D\geq6$, it is the repulsive component due to rotation that would be dominant for large $r$ asymptotically. It however turns out that for large values of rotation parameters, horizon occurs at $r/\mu<1$, and therefore  attractive term would again dominate over repulsive one closer to horizon. }

This interplay between attraction and repulsion gives rise to a distinctive dynamics for higher dimensional, $D\geq6$, rotating black holes. Very recently it had been explicitly shown \cite{Shaymatov20a} that six dimensional rotating black hole cannot be overspun for linear order accretion process. This is in contrast to what is generally true that at linear order it is always possible to overspin a black hole. The main purpose of this work is to show that is indeed true for in all dimensions $>6$ as well. That is, in all dimensions greater than five a rotating black hole having $n=[(D-1)/2]$ rotations cannot be overspun even under linear order accretion. Also note that even when overspinning is allowed at linear order, it is always overturned at non-linear order. Hence when it is not admitted at linear order, there is no question of it being overturned at non-linear order. Further a black hole having more than one rotation parameters and if one of which is zero, black hole has only one horizon and hence the question of overspinning does not arise. This is what we had first noticed for a five dimensional black hole with single rotation~\cite{Shaymatov19a}.

Thus in all dimensions greater than five, a rotating black hole would therefore always obey WCCC even for linear accretion.

The paper is organised as follows: In Sec.~\ref{Sec:metric} we briefly discuss general metric for higher dimensional rotating black hole in $D=2n+1,2n+2$ dimensions, which is followed by the analysis leading to the discussion of overspinning of black holes with $n-1$ and $n$ rotations in Sec.~\ref{Sec:III}. We end with a discussion in Sec.~\ref{Sec:Discussion}.

\section{Black holes in higher dimensions}\label{Sec:metric}

The line element of the higher dimensional rotating Myers-Perry black hole \cite{Myers-Perry86} is in odd $d=2n+1$ dimension given by
\begin{eqnarray}\label{2n+1}
ds^2&=&-dt^2 + (r^2+a^2_{i})\left(d\mu_{i}^2+\mu_{i}^2d\phi^2_{i}\right)\nonumber\\&+&\frac{\mu r^2}{\Pi F}\left(dt +a_{i}\mu_{i}^2d\phi_{i}\right) +\frac{\Pi F}{\Delta}dr^2\, ,
\end{eqnarray}
where
\begin{eqnarray}
\Delta = \Pi-\mu r^2,
\end{eqnarray}
and the metric in even $D=2n+2$ takes the form
\begin{eqnarray}\label{2n+2}
ds^2&=&-dt^2+r^2d\alpha^2 + (r^2+a^2_{i})\left(d\mu_{i}^2+\mu_{i}^2d\phi^2_{i}\right)\nonumber\\&+&\frac{\mu r}{\Pi F}\left(dt +a_{i}\mu_{i}^2d\phi_{i}\right) +\frac{\Pi F}{\Delta}dr^2\, ,
\end{eqnarray}
with
\begin{eqnarray}
\Delta = \Pi-\mu r.
\end{eqnarray}
Further we have
\begin{eqnarray}
F &=& 1-\frac{a_{i}^2\mu_{i}^2} {r^2+a_{i}^2}\, , \nonumber\\
\Pi &=&(r^2+a_1^2)...(r^2+a_i^2) \, .
%
\end{eqnarray}
Here $n= [(D-1)/2]$ is the maximum number of rotation parameters in dimension $D$, and $\mu$ and $a_i$ are respectively mass and rotation parameters. Note that $\mu_i$ are direction cosines satisfying $ \Sigma \mu_i^2=1$ and $ \Sigma \mu_i^2 + \alpha^2 =1$ for $D = 2n+1, 2n+2$, respectively.

Black hole horizon is given by $\Delta=0$ and which in odd and even dimensions will respectively read as follows:
\begin{eqnarray}
(r^2+a_1^2)...(r^2+a_i^2) - \mu r^2 =0\, ,
%
\end{eqnarray}

and
\begin{eqnarray}
(r^2+a_1^2)...(r^2+a_i^2) - \mu r =0 \, .
%
\end{eqnarray}

Looking at the above polynomial equations in the two cases, it is clear that in the former $D=2n+1$, it has two positive, two negative and rest all complex conjugate roots, while for the latter $D=2n+2$, there occur only two positive and rest all complex conjugate roots.

Thus black hole would always have two horizons and hence it could be overspun, whether that really happens or not is what we investigate in the following.

\section{Overspinning of black holes }\label{Sec:III}

It turns out that black hole with $(n-1)$ rotations behave characteristically differently from that with maximum allowed $n = [(D-1)/2]$ rotations in a given dimension $D$. We shall consider these two cases separately.

\subsection{$(n-1)$ rotations}

It turns out that when one of rotations is switched off (i.e. $(n-1)$ instead of $n$ rotations), black hole has only one horizon irrespective of dimension being odd $D=2n+1$ or even $D=2n+2$. This is because the horizon equation, $\Delta = 0$, Eq.~(\ref{Eq:delta}), has only one positive root and rest all complex conjugates for both odd and even $D=2n+1, 2n+2$ dimensions. Since there occurs only one horizon,  thereby there is no extremality condition defined and hence the question of overspinning doesn't arise. However to illustrate it with a specific example, let us consider seven dimensional black hole with two instead of maximum allowed three rotations. The horizon equation, $\Delta = 0$ takes the form
\begin{eqnarray}\label{Eq:delta}
\frac{(r^2+a^2)...(r^2+a_{n}^2)}{r^{2(n-1)}}-\mu r^{5-D}=0 \, ,
\end{eqnarray}
which will be of the form
\begin{eqnarray}\label{Eq:delta}
r^{2n} + f_1(a_i) r^{2n-2} + ... - \mu r^{3-D+2n} + a_1^2a_2^2 ...a_n^2 = 0 \, ,
\end{eqnarray}
with $f_{1}(a_{i})=a_1^2+a_2^2+...+a_{n}^2$.
For $D=7$ it solves to give
\begin{eqnarray}
r_{\pm}^2 = -\frac{a^2+b^2}{2}\pm\frac{1}{2}\sqrt{\left(a^2+b^2\right)^2 -4\mu}\, ,
\end{eqnarray}
where we have denoted the two rotations by $a, b$. This clearly shows that there is only one positive real root, and thus there exists only one horizon\footnote{This result that there exists no extremal solution for horizon when one or more rotations being
zero has also been noticed in Ref.~\cite{Doukas11} while considering Myers-Perry black hole in $D<10$. Here we have
given a direct and simple proof for any dimension $D$.}. This will be the case in all higher dimensional $D>4$ black holes with $(n-1)$ rotations.

The necessary condition for overspinning of a black hole is existence of two horizons so that extremality condition is defined to do further analysis. Since there exists only one horizon, the question of further investigation does not arise. That is, a black hole with one rotation less than the maximum allowed in a given dimension can never be overspun. Let us couch this general result as a theorem: \\

\textit{Theorem I: A black hole in a given dimension having  one of its rotations zero (i.e. $(n-1)$ rotations) can never be overspun and hence would always obey WCCC.}
\begin{figure*}
\centering
  \includegraphics[width=0.3\textwidth]{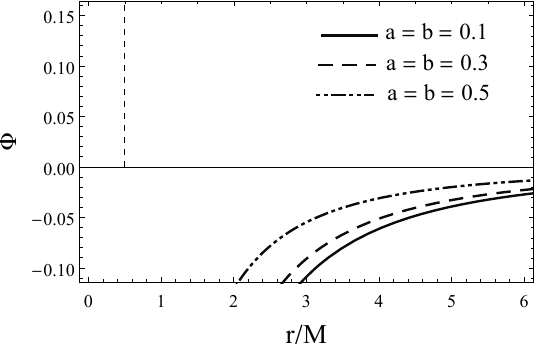}
  \includegraphics[width=0.3\textwidth]{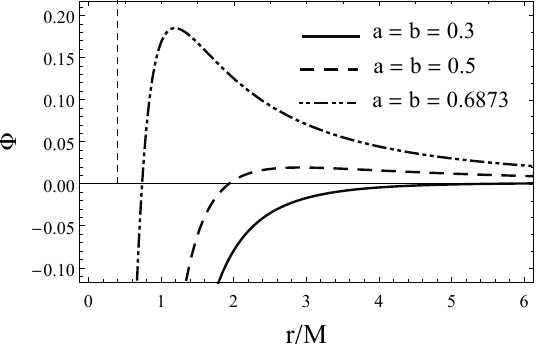}
  \includegraphics[width=0.3\textwidth]{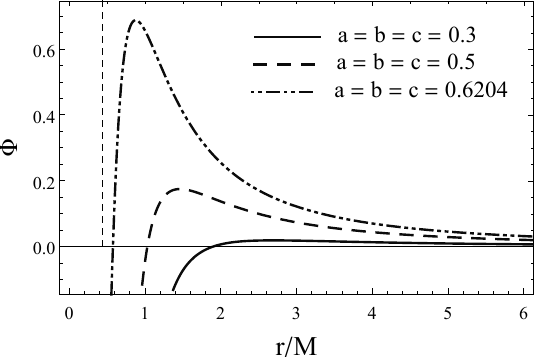}

\caption{\label{fig1}From left to right: Potential $\Phi(r)$ for $D=5, 6, 7$ is plotted against $r/M$. In all panels, vertical dashed line indicates the horizon for near extremal values of rotation parameters for which plot is shown by dot-dashed lines. }
\end{figure*}

\subsection{$n$-rotations}

As we have shown earlier that for the maximum allowed $n$ rotations in a given dimension $D$, the horizon equation $\Delta = 0$ always has two positive roots irrespective of $D = 2n+1, 2n+2$ indicating existence of two horizons for black hole. It thus satisfies the necessary condition for overspinning. However it has been shown by explicit calculation in \cite{Shaymatov20a} that six dimensional black hole with two rotations cannot be overspun under linear test particle accretion. Could this be the case in all higher dimensions as well? However we also know that a five dimensional black hole with two rotations can be overspun under linear accretion. That is,  there occurs a transition from overspinning to no overspinning as we go from five to six dimension. This is what we attempt to understand in the following.

What is the critical change that occurs while going from five to six dimension? We know that gravity becomes sharper with dimension as gravitational potential due to mass goes as $1/r^{D-3}$. Further there is also contribution to potential from rotation which is though repulsive, opposite in character to that due to mass. Gravitational potential in the leading order could be written as
$\Phi(r) = \Delta/r^2 -1$, which for the familiar four dimensional rotating Kerr black hole is $\Phi(r) = -M/r + a^2/r^2$. Clearly as $D$ increases, the first term will become sharper with $D-3$ while the second term remains unchanged. Note that increase in $D$ will entail increase in rotation parameters which would, apart from contribution to the second term, further contribute higher order terms with positive sign. All contributions due to rotations are repulsive, and the leading $1/r^2$ term will for large $r/M>1$ dominate over attractive $1/r^{D-3}$ for  $D>5$. In $D=5$, the leading order term is $-(\mu-a^2)/r^2$ which would remain attractive for $\mu>a^2$. This is the critical change that comes about when we go beyond five dimensions. In $D>5$, attractive component would decay faster than repulsive $1/r^2$ and hence at infinity resultant force would be repulsive. This is in contrast to usual asymptotically flat spacetimes and this transition occurs at $D=6$; i.e $D=5$ is the threshold beyond which overall gravitational force is repulsive at large $r$.

By recalling Eq.~(\ref{Eq:delta}) we write effective gravitational potential for black holes with $n$ rotations,
\begin{eqnarray}\label{Eq:phi}
\Phi(r)\approx\frac{\Delta}{r^2}-1=\frac{(r^2+a^2)...(r^2+a_{n}^2)}{r^{2n}}-\frac{\mu}{r^{D-3}}-1\, .
\end{eqnarray}

For a clearer understanding, let's write the above equation explicitly for $D= 5, 6, 7$ as follows:
\begin{eqnarray}
\Phi_{5D}(r)&=&-\frac{\mu}{r^2}+\frac{a^2+b^2}{r^2}+\frac{a^2b^2}{r^4}\, ,\\
\Phi_{6D}(r)&=&-\frac{\mu}{r^3}+\frac{a^2+b^2}{r^2}+\frac{a^2b^2}{r^4}\, ,\\
\Phi_{7D}(r)&=&-\frac{\mu}{r^4}+\frac{a^2+b^2+c^2}{r^2}+\frac{a^2b^2+b^2c^2+a^2c^2}{r^4}\nonumber\\&&+\frac{a^2b^2c^2}{r^6}\, .
\end{eqnarray}

Figs.~\ref{fig1} and \ref{fig2} respectively show plots of $\Phi(r)$ and its derivative from left to right for $D = 5, 6, 7$. This celarly shows that overall acceleration is attractive  all through for $D=5$ while it is repulsive for $D=6, 7$ for large $r/M$. This would be the same in all higher dimensions $\geq6$. It however turns attractive closer to horizon which is because horizon occurs for $r/M<1$ where attractive component, $1/r^{D-3}$ rides over repulsive $1/r^2$ as well as relative dominance of mass over rotation parameters.

In \cite{Shaymatov20a} it has been shown by explicit calculation that six dimensional black hole with two rotations cannot be overspun under linear accretion. As we have seen above that in all higher dimensions $>6$ gravitational dynamics would be similar to that in $D=6$, hence what happens in six dimensions should hold true in all higher dimensions as well. That is, black holes having the maximum number of allowed rotations in all $D\geq6$ cannot similarly be overspun. \\

We could thus state:
\textit{Theorem II: Black hole in dimension $>5$ can never be overspun under linear accretion and would thereby always obey WCCC .}

If a black hole cannot be overspun under linear accretion, it would continue to do so for non-linear perturbations because the latter always favours no overspinning and thereby WCCC.
\begin{figure*}
\centering
  \includegraphics[width=0.3\textwidth]{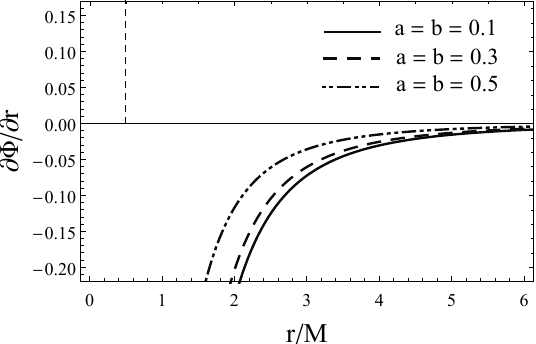}
  \includegraphics[width=0.3\textwidth]{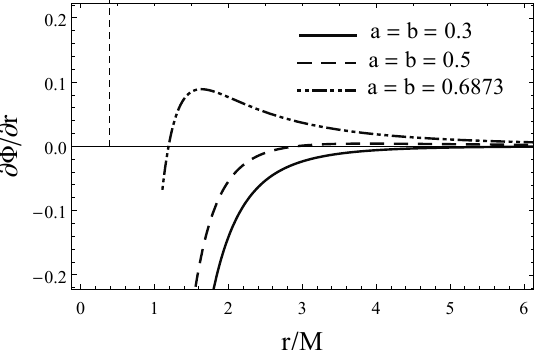}
  \includegraphics[width=0.3\textwidth]{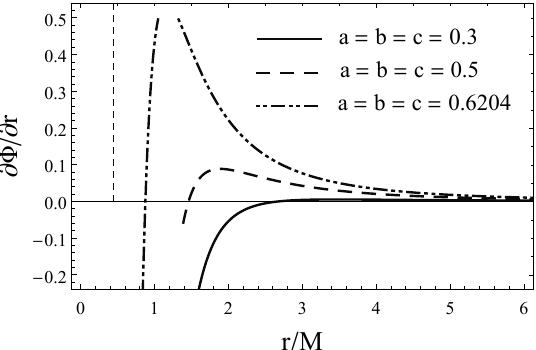}

\caption{\label{fig2} From left to right: $\partial \Phi(r)/\partial r$ for $D=5, 6, 7$ is plotted against $r/M$. In all panels, vertical dashed line indicates the horizon for near extremal values of rotation parameters for which plot is shown by dot-dashed lines. }
\end{figure*}

\section{Discussion}
\label{Sec:Discussion}

The interplay between attraction due to mass and repulsion due to rotation parameters of black hole gives rise to an interesting setting with richer dynamics. In $D\leq5$, the former dominates while the latter does for $D>5$ for large $r$. This is because potential due to mass goes as $1/r^{D-3}$ whereas that due to rotation in the leading order as $1/r^2$. The latter would dominate over the former in $D>5$ for large $r/M>1$. Thus five dimension is the upper threshold for overall gravity being attractive asymptotically.

A rotating black hole in dimension greater than or equal to six,  effective gravitational acceleration is repulsive at large $r/M>1$, however it turns attractive closer to horizon. This is due to the fact that horizon occurs at $r/M<1$ where attractive component picks up as well as relative dominance of mass over rotation parameters. Thus gravitational dynamics is characteristically different for rotating black hole in dimensions $<6$ and in $\geq6$. In the former case, attractive component is dominant all through while for the latter attraction is dominant only very close to horizon while repulsion dominates all through asymptotically. It is therefore natural to expect black hole to behave differently in the two dimensional range.

By an explicit calculation \cite{Shaymatov20a} it has been shown that six dimensional black hole cannot be overspun even under linear accretion. Since gravitational dynamics has the same character in all dimensions greater than six, hence whatever is true for six dimension should be true for in all dimensions greater than six. That means since six dimensional rotating black hole cannot be overspun, so would be the case for all higher dimensional black holes. No rotating black hole in dimensions greater than or equal to six could be overspun (\textit{Theorem II}). They would all therefore obey WCCC.

For overspinning of a black hole the necessary condition is existence of two horizons so that extremality condition is defined. Then further analysis could ensue to examine whether overspinning is admitted or not. If a black hole admits only one horizon, then the question of its overspinning does not arise.

In higher dimensions, black hole can have more than one rotation parameters, and maximum number of parameters in a given dimension is $n=[(D-1)/2]$. It turns out that if black hole has $(n-1)$ rotations instead of $n$, it admits only one horizon and hence it can never be overspun (\textit{Theorem I}). It would thus always obey WCCC.

These are the two main results of the paper which have been couched as the two theorems. In conclusion, we state that a rotating black hole in dimensions greater than or equal to six, or else one of its rotation parameters is switched off, always obeys the weak cosmic censorship conjecture.

It should be noted that a black hole with one horizon is altogether a different creature than the one with two. This is because its spacetime has radically different causal structure. The singularity at the center is spacelike in contrast the one with two horizons would have null or timelike singularity. The causal structure of the former is Schwarzschild-like while that of the latter is Kerr-like. \textit{Theorem I} thus indicates a very important physical property that distinguishes black holes having $(n-1)$ and $n$ rotations.

For \textit{Theorem II}, the key role is played by the critical property of transition of overall force from attractive to repulsive slightly away from horizon. This occurs at $D=6$; i.e. it is attractive for $D<6$ and repulsive when $D\geq6$. For $D=6$, WCCC has been proven by explicit calculations \cite{Shaymatov20a}. Since all higher dimensions $>6$ share this critical property, hence WCCC holds good for in all dimensions greater than six as well. This may not be as solid a proof as that of an explicit calculation, yet it has strong physical arguments and motivation. It is therefore not simply a speculation and guess work.

Finally we end up with an interesting and intriguing question, this analysis raises. Since overall gravity for a six and higher dimensional rotating black hole is repulsive for large $r$, how could such a black hole be formed by gravitational collapse of a cloud?\\

\section*{Acknowledgments}

S.S. acknowledges the support of the Uzbekistan Ministry for Innovative Development Projects No. VA-FA-F-2-008. and No. MRB-AN-2019-29.

\bibliographystyle{apsrev4-1}  
\bibliography{gravreferences}

 \end{document}